%% file: SDW.tex
\documentclass[prb,superscriptaddress,twocolumn]{revtex4-2}
\pdfoutput=1

\usepackage{mathtools}
\usepackage{newtxtext}
\usepackage[upint,varg]{newtxmath}
\usepackage{bm}
\usepackage{graphicx}
\usepackage{braket}
\usepackage{hyperref}
\usepackage[all]{hypcap}
\usepackage{tikz}

\newcommand*\diff{\mathop{}\!\mathrm{d}}

\renewcommand{\Re}{\operatorname{Re}}
\newcommand{\sgn}{\operatorname{sgn}}
\newcommand{\Tr}{\operatorname{Tr}}

\renewcommand{\Delta}{\varDelta}

\begin{document}
 
\title{First-order phase transition between superconductivity and charge/spin-density wave causes their coexistence in organic metals}

\author{Seidali\,S.~Seidov}
\affiliation{National University of Science and Technology ''MISiS'', 119049, Moscow, Russia}

\author{Vladislav\,D.~Kochev}
\email[e-mail:]{vd.kochev@misis.ru}
\affiliation{National University of Science and Technology ''MISiS'', 119049, Moscow, Russia}

\author{Pavel\,D.~Grigoriev}
\email[Corresponding author, e-mail:]{grigorev@itp.ac.ru}
\affiliation{L.\,D.~Landau Institute for Theoretical Physics, 142432, Chernogolovka, Russia}
\affiliation{National University of Science and Technology ''MISiS'', 119049, Moscow, Russia} 

%\date{\today }

\begin{abstract}
The interplay between superconductivity (SC) and spin/charge density wave (DW) 
in organic metals shows many similarities to high-$T_\text{c}$ superconductors. It also 
contains many puzzles, for example, the anisotropic SC onset observed and 
the severalfold increase of the upper critical field $H_\text{c2}$ in the coexistence region, 
as well as the microscopic origin of SC/DW phase separation there.
In this paper, by the direct calculation of the Landau
expansion for DW free energy, we argue that the phase transition between DW and metallic/SC
phase in organic superconductors goes by first order at low enough temperature, which explains
the spatial segregation of DW and SC at large length scale, consistent with
experimental observations. This first-order phase transition is not directly
related to SC and happens even above the SC transition temperature.
\end{abstract}

\maketitle

\section{Introduction}

The properties of metals with a charge-density-wave (CDW) or
spin-density-wave (SDW) ground state attract great attention since the
fifties (see, e.g., [\onlinecite{Gruner1994,MonceauAdvPhys}]). It often
competes with superconductivity (SC) \cite{Review1Gabovich,
ReviewGabovich2002,MonceauAdvPhys}, e.g., in most high-$T_\text{c}$ superconductors,
both cuprate \cite{XRayNatPhys2012, XRayPRL2013,
XRayCDWPRB2017,Tabis2014,Science2015Nd,Wen2019} and iron-based \cite%
{ReviewFePnictidesAbrahams, ReviewFePnictides2}, and remains a subject of
active investigation in many other materials, including transition metal
dichalcogenides \cite{CDWSCNbSe2, NbSe2NatComm}, organic superconductors
(OS) \cite{Ishiguro1998,AndreiLebed2008-04-23,Hc2Pressure,Vuletic,Kang2010,
ChaikinPRL2014, LeeBrownNMRDomains,
LeeTriplet,LeeBrownNMRTriplet,Gerasimenko2013,Gerasimenko2014,Yonezawa2018,CDWSC}
and even materials with nontrivial topology of band structures \cite{Yu2021}.

The OS are interesting for studying the interplay between CDW/SDW and SC
because their phase diagram, layered crystal structure and many other
features resemble those in cuprate and iron-based high-$T_\text{c}$ superconductors,
but they are simpler and more convenient for investigation. By changing the
chemical composition one can control the electronic dispersion in OS in a
wide interval. One can grow rather large and pure monocrystals of organic
metals, so that their electronic structure can be experimentally studied by
high-magnetic-field tools \cite{Kartsovnik2004Nov}. Finally, the OS are
simpler for theoretical study because of weaker electronic correlations \cite%
{Ishiguro1998, AndreiLebed2008-04-23}. Therefore, the OS are very helpful
for disentangling various factors affecting the electronic and
superconducting properties, which is hard to do in cuprates and other
strongly correlated materials. SC and DW coexist even in relatively weakly
correlated OS, such as (TMTSF)$_{2}$PF$_{6}$ \cite{LeeBrownNMRDomains,
Vuletic, Kang2010, ChaikinPRL2014}, (TMTSF)$_{2}$ClO$_{4}$ \cite%
{Gerasimenko2014, Yonezawa2018} or $\upalpha $-(BEDT-TTF)$_{2}$KHg(SCN)$_{4}$ 
\cite{CDWSC}. In these materials the density wave (DW) is suppressed by some
external parameter, such as pressure or cooling rate. Similar to cuprates,
the SC transition temperature $T_\text{cSC}$ is the highest in the coexistence
region near the quantum critical point where DW disappears. The upper
critical field $H_\text{c2}$ is often several times higher in the coexistence
region than in a pure SC phase \cite{Hc2Pressure,CDWSC}, the effect
potentially useful for applications.

The microscopic structure of SC and DW coexistence is important for
understanding the DW influence on SC properties and $T_\text{cSC}$. The DW and SC phase separation may happen in the
momentum or coordinate space. The first scenario assumes a spatially uniform
structure, when the Fermi surface (FS) is partially gapped by DW and the
ungapped parts give SC \cite{MonceauAdvPhys, GrigorievPRB2008}. It is 
similar to most other CDW materials \cite{MonceauAdvPhys,Review1Gabovich,ReviewGabovich2002}. 
The second scenario assumes that SC and DW phases are spatially separated on a
microscopic or macroscopic scale, depending on the ratio of SC domain size 
$d$ and the SC coherence length $\xi_ \text{SC}$. The temperature resistivity
hysteresis observed in (TMTSF)$_{2}$PF$_{6}$ \cite{Vuletic} supports the
spatial SC/DW separation. The width of SC transition increases with the 
increase of disorder, controlled by the cooling rate in (TMTSF)$_{2}$ClO$_{4}$, 
which also indicates a spatial SDW/SC segregation \cite{Yonezawa2018} similar to 
granular superconductors. The microscopic SC domains of width $d$ comparable to 
or even less than the SC coherence length $\xi_ \text{SC}$
may appear in the soliton DW structure, where SC emerges inside the soliton
walls \cite{BrazKirovaReview, SuReview, GrigPhysicaB2009, GG, GGPRB2007,
GrigPhysicaB2009}. However, the angular magnetoresistance oscillations
inside the parametric region of SC/DW coexistence observed in (TMTSF)$_{2}$PF%
$_{6}$ \cite{ChaikinPRL2014} and in (TMTSF)$_{2}$ClO$_{4}$ \cite%
{Gerasimenko2013} seem to be consistent with only a macroscopic spatial
phase separation with SC domain width $d>1$~\textmu m. The SC upper critical
field $H_\text{c2}$ may theoretically exceed several times the $H_\text{c2}$ without
DW coexistence in all the above scenarios \cite{GrigorievPRB2008,
GrigPhysicaB2009}, provided the SC domain width is smaller or comparable to the
penetration depth $\lambda$ of magnetic field to the superconductor.

The puzzling feature of SDW/SC coexistence in (TMTSF)$_{2}$PF$_{6}$, long
unexplained in any scenario, is the anisotropic SC onset \cite%
{Kang2010,ChaikinPRL2014}: with the increase in pressure at $P=P_\text{c2}\approx
6.7$~kbar the SC transition and the zero resistance is first observed only
along the least-conducting interlayer $z$-direction, then at $%
P=P_\text{c1}\approx 7.8$~kbar along $z$- and $y$-directions, and only at $%
P=P_\text{c0}\approx 8.6$~kbar in all directions, including the most conducting $%
x $-direction. This is opposite to a weak intrinsic interlayer Josephson
coupling, typical in high-$T_\text{c}$ superconductors \cite{Tinkham}. Other
organic metals manifest similar anisotropic SC onset in the region of
coexistence with DW \cite{Gerasimenko2014}. The observed \cite{Kang2010,
ChaikinPRL2014,Gerasimenko2014} anisotropic zero-resistance transition temperature 
$T_\text{cSC}$ seems to contradicts the
general rule that the percolation threshold in macroscopic heterogeneous media
must be isotropic \cite{PercolationEfros}, provided the SC inclusions are
not thin filaments \cite{Kang2010} connecting opposite edges of a sample.
However, such a filament scenario cannot be substantiated microscopically, neither in
(TMTSF)$_{2}$PF$_{6}$ nor in (TMTSF)$_{2}$ClO$_{4}$. 
This paradox was resolved recently \cite{Kochev2021,KesharpuCrystals2021} by
assuming a spatial SC/DW separation and studying the percolation in
finite-size samples of the thin elongated shape, relevant to the experiments
in (TMTSF)$_{2}$PF$_{6}$ \cite{Kang2010,ChaikinPRL2014} and (TMTSF)$_{2}$ClO$%
_{4}$ \cite{Gerasimenko2014, Yonezawa2018}. Similar effect was observed and used
to study the SC domain shape and size in other superconductors, for example, 
in FeSe \cite{Sinchenko2017,Grigoriev2017,Grigoriev2023FeSe} or other organic 
metals \cite{Seidov2018}. This supports the scenario of
spatial SC/DW segregation in a form of rather large domains of width $d>1$ \textmu m 
in organic superconductors. However, the microscopic reason for such phase 
segregation remains unknown and is the main goal of our study.

In this paper we argue that the phase transition between DW and metallic\slash SC
phase in OS goes by first order at low enough temperature, which explains
the spatial segregation of DW and SC at large length scale, consistent with
experimental observations. This first-order phase transition is not directly
related to SC and happens even at $T>T_\text{cSC}$ because of the suppression of DW
by the deterioration of FS nesting, which is controlled by pressure in
(TMTSF)$_{2}$PF$_{6}$ and in $\upalpha$-(BEDT-TTF)$_{2}$KHg(SCN)$_{4}$, or by
cooling rate affecting the anion ordering in (TMTSF)$_{2}$ClO$_{4}$. In Sec.
\ref{sec:model} we describe the driving parameter of DW-metal phase transition 
in various quasi-1D organic metals. In Sec. \ref{sec:mean-field} formulate 
the mean-field approach for the free energy in the DW state and write down 
its Landau expansion.  In Sec. \ref{sec:coefficients} we perform the explicit 
calculation of the DW free-energy expansion for the model described in Sec.
\ref{sec:model} and analyze the DW phase diagram of organic superconductors.
In particular, we study the range of temperature and electron-spectrum parameters 
%of band splitting and quasi-1D electron dispersion 
where the phase transition 
from DW to metallic state is of first order.
In Sec. \ref{sec:discussion}  discuss our results in connection with the 
experimental observations in OS. 

\section{The model: driving parameters of phase transition to density wave}\label{sec:model}

\subsection{Quasi-one-dimensional electron dispersion and pressure as a driving
parameter of DW-metal/SC phase transition in organic superconductors}

In Q1D organic metals \cite{Ishiguro1998, AndreiLebed2008-04-23}, at which
our present study is mainly aimed, the free electron dispersion near the
Fermi level without magnetic field is approximately given by 
\begin{equation}
\varepsilon ({ \bm{k}})=\hbar v_\text{F}(|k_{x}|-k_\text{F})+t_{\perp }({%
 \bm{k}}_{\perp }),  \label{dispersion1}
\end{equation}%
where $v_\text{F}$ and $k_\text{F}$ are the Fermi velocity and Fermi momentum in the
chain $x$-direction. We consider a quasi-1D metal with dispersion (\ref%
{dispersion1}) where the function $t_{\perp }({ \bm{k}}_{\perp })$ is
given by the tight-binding model: 
%\begin{equation}
%\begin{split}
%t_{\perp }({ \bm{k}}_{\perp }) = &-2t_{b}\cos (k_{y}b)-2t_{b}^{\prime
%}\cos (2k_{y}b) - \\
%& - 2t_{c}\cos (k_{z}c),
%\end{split}
%\label{dispersion}
%\end{equation}
\begin{equation}
	\begin{split}
		t_{\perp }({ \bm{k}}_{\perp }) &= -2  t_{b}\cos (k_{y}b) - 2t_{b}^{\prime} \cos (2k_{y}b)\\ 
		&- 2 t_{c}\cos (k_{z}c),
	\end{split}
	\label{dispersion}
\end{equation}%
where $b$ and $c$ are the lattice constants in $y$- and $z$-directions
respectively. The corresponding Fermi surface (FS) of quasi-1D metals
consists of two slightly warped sheets separated by $2k_\text{F}$ and roughly
possesses the nesting property. It leads to the Peierls instability and
favors the formation of CDW or SDW at low temperature, which competes with
superconductivity. In the metallic phase the corresponding density of
electron states at the Fermi level per one spin components per unit length $%
L_{x}$ of one chain is $\nu_\text{F} = 1/ (\pi \hbar v_\text{F})$.

The dispersion along the $z$-axis is assumed to be much weaker than the
dispersion in the $y$-direction. Therefore we omit the second harmonic $%
\propto \cos (2k_{z}c)$ in the dispersion relation (\ref{dispersion}).
Since the terms $\propto \cos (k_{z}c)$ and $\propto \cos (k_{y}b)$ do not
violate the perfect nesting condition 
\begin{equation}
\varepsilon ( \bm{k}+ \bm{Q}_0)=-\varepsilon ( \bm{k}),
\label{2}
\end{equation}%
they do not influence the physics discussed below unless the nesting vector
become shifted in the $y$-$z$ plane. We do not consider such a shift in the
present study. Hence, only the ''antinesting'' parameter $t_{b}^{\prime }\sim
t_{b}^{2}/(v_\text{F}k_\text{F})$ is important for the DW phase diagram. The electron
dispersion along the $z$-axis is ignored below.

With the increase of applied pressure $P$ the lattice constants decrease.
This enhances the interchain electron tunneling and the transfer integrals.
The increase of $t_{b}^{\prime } (P)$ with pressure deteriorates the
FS nesting and decreases the DW transition temperature $T_ \text{cDW}(P)$ 
\cite{Ishiguro1998,AndreiLebed2008-04-23}. 
There is a critical pressure $P_\text{c}$ and the corresponding
critical value $t_{b}^{\prime \ast }=t_{b}^{\prime } (P_\text{c}) $ at
which $T_ \text{cDW} ( P_\text{c}) =0$ and one has a quantum critical point. 
The electronic properties at this DW quantum critical point are additionally complicated
by superconductivity emerging at $T<T_ \text{cSC}$ at $P>P_\text{c}$. In organic metals
SC appears even earlier at pressure $P>P_\text{c1}$, where $P_\text{c1}<P_\text{c}$, and 
there is a finite region $P_\text{c1}<P<P_\text{c}$ of SC-DW coexistence 
\cite{Kang2010, ChaikinPRL2014,CDWSC}.
%(see Fig. \ref{FigPhDia})
This simple model qualitatively describes the
phase diagram observed in (TMTSF)$_{2}$PF$_{6}$ and $\upalpha $-(BEDT-TTF)$%
_{2} $KHg(SCN)$_{4}$.
In (TMTSF)$_{2}$PF$_{6}$ the resistivity measurements in a magnetic field 
give \cite{Chaikin1996} $t_{b}^{\prime \ast } \approx 11.3\pm 0.2$~K, while at ambient pressure  
$t_{b}^{\prime } \approx 4.5\pm 0.3$~K and $T_\text{cDW}=12$~K. 

\subsection{Other driving parameters of DW-metal phase transition}

\begin{figure}[tb]
	\centering\includegraphics{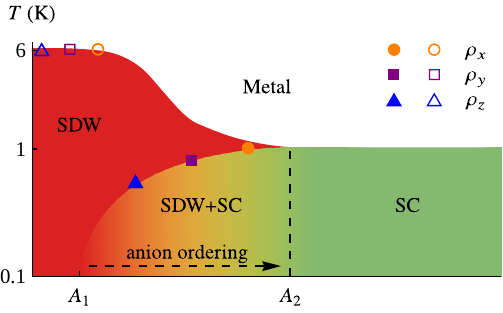}
	\caption{Phase diagram of SDW in (TMTSF)$_{2}$ClO$_{4}$ recreated from resistivity data in Ref. \cite{Gerasimenko2014}.}
	\label{FigPhDiaClO4}
\end{figure}

In (TMTSF)$_{2}$ClO$_{4}$ there is an additional degree of freedom and
driving parameter. The ClO$_{4}$ anions, which lack the inversion symmetry
and possess an electric dipole moment, order below the transition
temperature $T_{  \text{AO}}=24$~K $>T_{\text{cDW}}$ with wave vector $%
 \bm{Q}_{  \text{AO}}=(0,\pi/b,0)$ \cite%
{Ishiguro1998,AndreiLebed2008-04-23}. This anion ordering (AO) doubles the
lattice along the $y$-axis and splits the FS in four open sheets (see, e.g.,
Fig. 1b of Ref. \cite{Gerasimenko2013} or Fig. 3b and 5b of Ref. \cite%
{Alemany2014}). The latter deteriorates FS nesting and suppresses DW \cite%
{AndreiLebed2008-04-23,Takahashi1982,Sengupta2001,Alemany2014,Guster2020},
giving rise to superconductivity \cite%
{AndreiLebed2008-04-23,Gerasimenko2013,Gerasimenko2014,Yonezawa2018,Aizawa2018}%
. 

The electron dispersion in (TMTSF)$_{2}$ClO$_{4}$ in the anion-ordered phase
in the new folded Brilluin zone can also be approximately described by Eq. (\ref{dispersion1}) with two
branches, splitted by $2\Delta _{  \text{AO}}$ (see Appendix \ref{sec:AppA}): %\begin{equation}
\begin{equation}
	t_{\perp}( \bm{k}_{\perp }) \simeq \pm 
	\left[ \Delta _{\text{AO}} -2t_{b}^{\prime }\cos (2k_{y}b)\right] .  \label{tAO}
\end{equation}%
This dispersion results to the antinesting term given by Eq. (\ref{epAO}), which controls the 
SDW phase diagram in (TMTSF)$_{2}$ClO$_{4}$. 

The anion ordering can be controlled by the cooling rate near $T_ \text{AO}=24$~K.
In rapidly cooled (quenched) samples the anions remain disordered, and the
DW is not suppressed, while SC does not appear. The slowly cooled (relaxed)
samples have $\Delta _ \text{AO}\approx 14$~meV \cite{Alemany2014,Aizawa2018}, which destroys
the SDW and leads to a homogeneous superconductivity. 
 Hence, the cooling
rate in (TMTSF)$_{2}$ClO$_{4}$ enables to reveal a rich phase diagram even
at ambient pressure, where DW phase is favored for strong disorders (rapid
cooling), SC and DW coexist at intermediate, and a uniform SC sets in for
weak disorders (slow cooling). The phase diagram in (TMTSF)$_{2}$ClO$_{4}$
is qualitatively similar to that in (TMTSF)$_{2}$PF$_{6}$ with the
replacement of pressure by anion ordering, as illustrated in Fig. \ref{FigPhDiaClO4}. 

An external magnetic field $B$ via the Zeeman splitting energy 
$\Delta_ \text{Z}=\mu_\text{B} g B$ shifts in opposite directions the FS for different 
spin projections, which damps CDW but not SDW. The magnetic field also affects 
the orbital electron motion, which leads to the beautiful effect of field-
induced spin-density wave \cite{Ishiguro1998, AndreiLebed2008-04-23}, to the 
quantized nesting and the complicated phase diagram. However, if the magnetic 
field is directed along the conducting layers, its orbital effect is weak, and 
the new dispersion is given by 
\begin{equation}
	t_{\perp }({ \bm{k}}_{\perp }) 
	\simeq -2t_{b}\cos(k_{y}b) -2t_{b}^{\prime }\cos (2k_{y}b) \pm \Delta_ \text{Z}.
	\label{tZ}
\end{equation}
This results to the same antinesting term (\ref{epAO}) as for AO, with the 
replacement $\Delta_ \text{AO} \to \Delta_ \text{Z}$. The CDW phase diagram in a 
magnetic field acting only by the Zeeman splitting is already quite interesting 
and complicated \cite{Grigoriev2005}. 

\section{Mean field approach and Landau expansion for the density wave}\label{sec:mean-field}

\subsection{Bogoliubov transformation}

The electron Hamiltonian consist of the free electron part $H_{0}$ and
the interaction part $H_{\text{int}}$: 
\begin{equation}
\begin{split}
& H=H_{0}+H_{\text{int}}, \\
& H_{0} = \sum_{\bm{k}\sigma} \varepsilon(\bm{k}) a_{\bm{k}\sigma}^{\dagger} a_{\bm{k}\sigma}, \\
& H_{\text{int}} = \frac{1}{2} 
\sum_{\substack{\bm{k}\bm{k}^{\prime}\bm{\mathcal{Q}} \\ \sigma\sigma'}} V_{\bm{\mathcal{Q}}\sigma\sigma'} 
a_{\bm{k}+\bm{\mathcal{Q}} \sigma}^{\dagger} a_{\bm{k}^{\prime}-\bm{\mathcal{Q}} \sigma'}^{\dagger}
a_{\bm{k}^{\prime} \sigma'} a_{\bm{k} \sigma}.
\end{split}
\label{H}
\end{equation}%
The coupling function $V_{\sigma\sigma'}( \bm{\mathcal{Q}})$ contains all types of electron-electron 
($e$-$e$) interaction: Coulomb repulsion, phonon-mediated attraction, etc. For 
the usual Coulomb or phonon-mediated $e$-$e$ interaction 
$V_{\sigma\sigma'}( \bm{\mathcal{Q}})$ is independent on the indices $\sigma ,\sigma'$. 
In the Hubbard model the dependence of $e$-$e$ interaction on spin indices 
appears because two electrons with the same spin orientation cannot occupy the 
same quantum state on the same site due to the Pauli principle, and therefore do 
no have the strong on-site $e$-$e$ interaction. The strong Coulomb repulsion 
is weakened by the metallic screening, therefore the DW order parameter is much 
smaller than the bandwidth or the Fermi energy in most metals, including organic 
superconductors. 

We consider the interactions at the wave vectors 
$\bm{\mathcal{Q}} = \pm \bm{Q} = \pm (\bm{Q}_0 + \bm{q})$ close to
the nesting vector $\bm{Q}_0 = (2k_\text{F},\pi /b,\pi /c)$ with $|\bm{q}| \ll k_\text{F}$.
If the deviations from $\pm\bm{Q}_{0}$ are small, we can approximate the
interaction function as 
$V_{\sigma\sigma'}( \bm{\mathcal{Q}})\approx V_{\sigma\sigma'}( \bm{Q}_{0})= U$, so
\begin{equation}
	H_{\text{int}} =  
	U \sum_{\bm{k}\bm{k}^{\prime}\bm{\mathcal{Q}}}  
	a_{\bm{k}+\bm{\mathcal{Q}} \uparrow}^{\dagger} a_{\bm{k}^{\prime}-\bm{\mathcal{Q}} \downarrow}^{\dagger}
	a_{\bm{k}^{\prime} \downarrow} a_{\bm{k} \uparrow}.
\end{equation}
Next, in the mean-field approximation we decouple the four-operator term
\begin{multline}
	a_{\bm{k}+\bm{\mathcal{Q}} \uparrow}^{\dagger} 
	a_{\bm{k}^{\prime}-\bm{\mathcal{Q}} \downarrow}^{\dagger} 
	a_{\bm{k}^{\prime} \downarrow} a_{\bm{k} \uparrow} 
	\approx 
	\left\langle a_{\bm{k}^{\prime}-\bm{\mathcal{Q}} \downarrow}^{\dagger} 
	a_{\bm{k}^{\prime} \downarrow}\right\rangle 
	a_{\bm{k}+\bm{\mathcal{Q}} \uparrow}^{\dagger} 
	a_{\bm{k} \uparrow} \\
	+ 
	\left\langle a_{\bm{k}+\bm{\mathcal{Q}} \uparrow}^{\dagger} 
	a_{\bm{k} \uparrow} \right\rangle 
	a_{\bm{k}^{\prime}-\bm{\mathcal{Q}} \downarrow}^{\dagger} 
	a_{\bm{k}^{\prime} \downarrow} 
	- 
	\left\langle a_{\bm{k}^{\prime}-\bm{\mathcal{Q}} \downarrow}^{\dagger} 
	a_{\bm{k}^{\prime} \downarrow}\right\rangle
	\left\langle a_{\bm{k}+\bm{\mathcal{Q}} \uparrow}^{\dagger} 
	a_{\bm{k} \uparrow}\right\rangle,
\end{multline}
so
\begin{multline}
	H_{\text{int}} = U \sum_{\bm{k}\bm{k}^{\prime}} \Big[ \\
	\left\langle a_{\bm{k}^{\prime}-\bm{Q} \downarrow}^{\dagger} 
	a_{\bm{k}^{\prime} \downarrow}\right\rangle 
	a_{\bm{k}+\bm{Q} \uparrow}^{\dagger} 
	a_{\bm{k} \uparrow}
	+ 
	\left\langle a_{\bm{k}+\bm{Q} \uparrow}^{\dagger} 
	a_{\bm{k} \uparrow} \right\rangle 
	a_{\bm{k}^{\prime}-\bm{Q} \downarrow}^{\dagger} 
	a_{\bm{k}^{\prime} \downarrow}
	\\
	+
	\left\langle a_{\bm{k}^{\prime}+\bm{Q} \downarrow}^{\dagger} 
	a_{\bm{k}^{\prime} \downarrow}\right\rangle 
	a_{\bm{k}-\bm{Q} \uparrow}^{\dagger} 
	a_{\bm{k} \uparrow}
	+ 
	\left\langle a_{\bm{k}-\bm{Q} \uparrow}^{\dagger} 
	a_{\bm{k} \uparrow} \right\rangle 
	a_{\bm{k}^{\prime}+\bm{Q} \downarrow}^{\dagger} 
	a_{\bm{k}^{\prime} \downarrow}
	\\ 
	- 
	\left\langle a_{\bm{k}^{\prime}-\bm{Q} \downarrow}^{\dagger} 
	a_{\bm{k}^{\prime} \downarrow}\right\rangle
	\left\langle a_{\bm{k}+\bm{Q} \uparrow}^{\dagger} 
	a_{\bm{k} \uparrow}\right\rangle
	- 
	\left\langle a_{\bm{k}^{\prime}+\bm{Q} \downarrow}^{\dagger} 
	a_{\bm{k}^{\prime} \downarrow}\right\rangle
	\left\langle a_{\bm{k}-\bm{Q} \uparrow}^{\dagger} 
	a_{\bm{k} \uparrow}\right\rangle
	\Big].
\end{multline}
%
%We now set $\Delta _{ \bm{k}}=\Delta \delta _{ \bm{k},\bm{Q}}$,
%

We introduce the order parameter 
\begin{equation}  \label{eq:Delta_def}
	\Delta_{ \bm{Q}} 
	= -\frac{|U|}{2} \sum_{\bm{k} \sigma} \braket{a_{\bm{k}-{\bm{Q}}\sigma}^\dagger 
	a_{\bm{k}\sigma'}} = \Delta,  \quad \Delta_{-\bm{Q}} = \Delta^*.
\end{equation}
%The operators $a^\dagger_{k+Q}$ and $a_k$ in Eqs. (\ref{H})-(\ref{eq:H_MF}) correspond to the same spin 
%component for the CDW, and to different spin components for the SDW. 
In this expression and in Eq. (\ref{eq:H_MF}) below $\sigma' = \sigma$ and $U = -|U| < 0$ 
for the CDW (i.e. the operators correspond to electrons with the same spin component), 
and $\sigma' = -\sigma$ and $U = |U| > 0$ for the SDW (the spin components are different). This 
corresponds to the choice of spin $z$-axis parallel to the SDW polarization vector 
$\bm{l}$. In our case this does not reduce the generality of our model, because 
there are no any other special spin directions, as in the presence of magnetic 
field or spin-orbit interaction.
For a general orientation of the SDW 
polarization vector $\bm{l}$ one should use Eq. (10) of 
Ref. \cite{GrigorievPRB2008}.

Between the CDW and SDW, the system chooses a 
state with the highest transition temperature, which in our model corresponds to 
the largest coupling constant $U$, 
giving the greatest order parameter $\Delta$ and energy gain. 
%Note that the coupling function $V_{\sigma\sigma'}( \bm{Q})$ contains all types of e-e interaction: Coulomb repulsion, phonon-mediated attraction, etc. The strong Coulomb repulsion is, however, weakened by the metallic screening, therefore the DW order parameter is much smaller than the bandwidth or Fermi energy in organic superconductors. 
Below the transition temperature the CDW and SDW are coupled only in a magnetic 
field or another spin-dependent external perturbation \cite{Grigoriev2005}.  
Similarly to the choice between CDW and SDW, 
after the system chooses the optimal DW wave vector $\bm{Q}$, the interaction 
$V( \bm{Q})$ at other wave vectors $\bm{Q}$ does not affect the electronic states 
in the mean-field approximation \cite{Grigoriev2005}. 
 
For CDW definition ($\ref{eq:Delta_def}$) directly leads to
\begin{equation} \label{DeltaCDW}
	\Delta 
	= U \sum_{\bm{k}} \braket{
		a_{\bm{k}-{\bm{Q}} \downarrow}^\dagger a_{\bm{k} \downarrow}}
	= U \sum_{\bm{k}} \braket{
		a_{\bm{k}-{\bm{Q}} \uparrow}^\dagger a_{\bm{k} \uparrow}}
\end{equation}
and
\begin{equation} \label{intCDW}
	H_\text{int} = \sum_{\bm{k}\sigma} \left(
		\Delta^* a_{\bm{k}-\bm{Q} \sigma}^\dagger a_{\bm{k} \sigma}
		+ \Delta a_{\bm{k} \sigma}^\dagger a_{\bm{k}-\bm{Q} \sigma}
	 \right)
	 + \frac{2 |\Delta|^2}{|U|},
\end{equation}
which is consistent with Ref. \cite{Solyom_Fundamentals_3}.
For SDW analogous relations to (\ref{DeltaCDW})-(\ref{intCDW}) is
\begin{equation} \label{DeltaSDW}
	\Delta 
	= U \sum_{\bm{k}} \braket{
		a_{\bm{k}-{\bm{Q}} \downarrow}^\dagger a_{\bm{k} \downarrow}}
	= -U \sum_{\bm{k}} \braket{
		a_{\bm{k}-{\bm{Q}} \uparrow}^\dagger a_{\bm{k} \uparrow}}.
\end{equation}
and
\begin{multline}
	H_\text{int} = \sum_{\bm{k}} \Big(
		\Delta^* a_{\bm{k}-\bm{Q} \uparrow}^\dagger a_{\bm{k} \uparrow}
		+ \Delta a_{\bm{k} \uparrow}^\dagger a_{\bm{k}-\bm{Q} \uparrow} \\
		- \Delta^* a_{\bm{k}-\bm{Q} \downarrow}^\dagger a_{\bm{k} \downarrow}
		- \Delta a_{\bm{k} \downarrow}^\dagger a_{\bm{k}-\bm{Q} \downarrow}
	 \Big)
	 + \frac{2 |\Delta|^2}{U},
\end{multline}
accordingly to Ref. \cite{Solyom_Fundamentals_3}.
To check that Eq. (\ref{DeltaSDW}) reduces to Eq. (\ref{eq:Delta_def}), one
can make linear transformation $a_{\bm{k}\uparrow} = (c_{\bm{k}+} + c_{\bm{k}-})/\sqrt{2}$,
$a_{\bm{k}\downarrow} = (c_{\bm{k}+} - c_{\bm{k}-})/\sqrt{2}$.
We consider the equality 
$\braket{a_{\bm{k}}^\dagger a_{\bm{k}-\bm{Q}}} 
= \braket{a_{\bm{k}\uparrow}^\dagger a_{\bm{k} -\bm{Q} \uparrow}} 
= \braket{a_{\bm{k}\downarrow}^\dagger a_{\bm{k}-\bm{Q} \downarrow}}
$ for CDW and $
\braket{a_{\bm{k}}^\dagger a_{\bm{k}-\bm{Q}}} 
= \braket{a_{\bm{k} \uparrow}^\dagger a_{\bm{k}-\bm{Q} \downarrow}} 
= \braket{a_{\bm{k} \downarrow}^\dagger a_{\bm{k}-\bm{Q} \uparrow}}
$ for SDW.

Both Hamiltonians can be expressed in similar way as 
\begin{equation}  \label{eq:H_MF}
H_\text{MF} 
= \!\sum_{\bm{k}\sigma} \varepsilon(\bm{k}) a_{\bm{k}\sigma}^{\dagger} a_{\bm{k}\sigma} 
+ \sum_{\bm{k}\sigma} \left( \Delta a_{\bm{k}\sigma}^{\dagger} a_{\bm{k}-\bm{Q},\sigma'} + \operatorname{H.c.}\right) 
+ \frac{2|\Delta|^2}{|U|},
\end{equation}
where $\Delta$ defined in Eq. (\ref{eq:Delta_def}) and sum over spins defined
in the sense explained below it. This Hamiltonian can be diagonalized via 
Bogoliubov transformation. In order to do so we
shift $\bm{k} \to \bm{k}+\bm{Q}$ under the sum and
rewrite the Hamiltonian in a matrix form as 
\begin{equation}
\begin{split}
&H_\text{MF} = \sum_{ \bm{k}}' \psi^\dagger_{ \bm{k}} h_{%
 \bm{k}} \psi_{ \bm{k}} + \frac{2|\Delta|^2}{|U|} , \\
&h = 
\begin{pmatrix}
\varepsilon( \bm{k}) & \Delta^* \\ 
\Delta & \varepsilon( \bm{k} +  \bm{Q})%
\end{pmatrix} ,
\\
&\psi^\dagger_{\bm{k}} = 
\begin{pmatrix}
a^\dagger_{ \bm{k}} & a^\dagger_{ \bm{k} +  \bm{Q}}%
\end{pmatrix}%
,
\end{split}%
\label{HMF}
\end{equation}
where prime indicates the $1/2$ factor due to avoid double counting 
(or the summation over the reduced Brillouin zone in the case of 
commensurate DW). Leaving out the spin indices, we cancel the above mentioned 
factor, so we omit the prime. We also fix gauge of $\Delta$, considering it real, 
and drop the modulus sign from $|U|$.
The spectrum of the Hamiltonian is then given by the eigenvalues $E_\pm (%
 \bm{k})$ of the matrix $h$ which are 
\begin{equation}  \label{eq:Spectrum}
E_\pm( \bm{k}) = \varepsilon_+( \bm{k}) \pm \sqrt{%
\varepsilon_-^2( \bm{k}) + \Delta^2} ,
\end{equation}
where 
\begin{equation}  \label{epm}
\varepsilon_\pm ( \bm{k}) \equiv \frac{\varepsilon( \bm{k}) \pm
\varepsilon( \bm{k} + \bm{Q})}{2}.
\end{equation}
The operators $\gamma_{ \bm{k}}$, $\mu_{ \bm{k}}$ in which
the Hamiltonian is diagonal are defined by the eigenvectors of $h$. They are
obtained by a unitary transformation 
\begin{equation}
\begin{pmatrix}
\gamma_{ \bm{k}} \\ 
\mu_{ \bm{k}}%
\end{pmatrix}
= 
\begin{pmatrix}
u & v \\ 
-v & u%
\end{pmatrix}
\begin{pmatrix}
a_{ \bm{k}} \\ 
a_{ \bm{k} +  \bm{Q}}%
\end{pmatrix}%
\end{equation}
such that $u^2 + v^2 = 1$ and $\Delta (u^2 - v^2) + 2 u v \varepsilon_-(%
 \bm{k}) = 0$. Finally the diagonalized Hamiltonian is 
\begin{equation}
\begin{split}
H &= \sum_{ \bm{k}} \left[ E_+( \bm{k}) \gamma^\dagger_{ \bm{k%
}} \gamma_{ \bm{k}} + E_-( \bm{k}) \mu^\dagger_{ \bm{k}%
} \mu_{ \bm{k}} \right] \\
& - \frac{1}{2} \sum_{ \bm{k}} \left[ E_+( \bm{k}) - E_-( \bm{%
k}) \right] + \frac{2\Delta^2}{U}.
\end{split}%
\end{equation}

Now given that the Hamiltonian is diagonal in fermionic operators $\gamma _{%
 \bm{k}}$, $\mu _{ \bm{k}}$ and there are only four states $%
|n_{\gamma }n_{\mu }\rangle =\{|00\rangle ,|01\rangle ,|10\rangle
,|11\rangle \}$, the partition function $Z=\Tr\exp \{-H\mathrm{_{%
\text{MF}}}/T\}$ can be easily calculated:%
\begin{equation}
\begin{split}
Z& =e^{-2\Delta ^{2}/U}\prod_{ \bm{k}}e^{(E_{+}+E_{-})/2T} \\
& \times \Tr%
\begin{pmatrix}
1 & 0 & 0 & 0 \\ 
0 & e^{-E_{+}/T} & 0 & 0 \\ 
0 & 0 & e^{-E_{-}/T} & 0 \\ 
0 & 0 & 0 & e^{-(E_{+}+E_{-})/T}%
\end{pmatrix}\\
& =e^{-2\Delta ^{2}/U}\prod_{ \bm{k}}4\cosh \frac{E_{+}( \bm{k}%
)}{2T}\cosh \frac{E_{-}( \bm{k})}{2T}.
\end{split}%
\end{equation}%
And the free energy $F=-T\ln Z$ is 
\begin{equation}
F=-T\sum_{ \bm{k}}\ln \left( 4\cosh \frac{E_{+}( \bm{k})}{2T}%
\cosh \frac{E_{-}( \bm{k})}{2T}\right) +\frac{2\Delta ^{2}}{U}.
\label{eq:F}
\end{equation}

\subsection{Free energy expansion}

We are interested in the Landau-Ginzburg expansion coefficients of the free
energy in powers of $\Delta $: 
\begin{equation}
F \simeq \frac{A}{2}\Delta ^{2}+\frac{B}{4}\Delta ^{4}+\ldots
\label{eq:F_expansion}
\end{equation}%
By differentiating this expression one obtains 
\begin{equation}
\frac{\partial F}{\partial \Delta } \simeq A\Delta +B\Delta ^{3}+\ldots
\label{eq:dF_expansion}
\end{equation}%
This means that the free energy expansion coefficients can be calculated by
expanding into the Taylor series not the free energy itself, but the
function 
\begin{equation}
\begin{split}
& F^{\prime }=\frac{\partial F}{\partial \Delta }=- \sum_{ \bm{k}}\frac{\Delta }{2\sqrt{%
\varepsilon _{-}^{2}( \bm{k})+\Delta ^{2}}} \\
& \times \left( \tanh \dfrac{E_{+}( \bm{k})}{2T}%
-\tanh \dfrac{E_{-}( \bm{k})}{2T}\right) +\frac{4\Delta }{U}.
\end{split}
\label{eq:dF}
\end{equation}%
Next we use the following relation to represent the hyperbolic tangents as a
sum over odd Matsubara frequencies $\omega =\pi T(2n+1)$: 
\begin{equation}
 \frac{1}{4\alpha }\left\{ \tanh \frac{\alpha -\beta }{2T}+\tanh \frac{%
\alpha +\beta }{2T}\right\}
 =T\sum_{\omega }\frac{1}{(\omega +i\beta )^{2}+\alpha ^{2}}.
\end{equation}%
Substituting $\alpha =\sqrt{\varepsilon _{-}^{2}({ \bm{k}})+\Delta
^{2}}$ and $\beta =\varepsilon _{+}({ \bm{k}})$ we find 
\begin{equation}
\begin{split}
& T\sum_{\omega }\frac{1}{[\omega +i\varepsilon _{+}({ \bm{k}}%
)]^{2}+\varepsilon _{-}^{2}({ \bm{k}})+\Delta ^{2}} \\
& =\frac{\tanh \dfrac{E_{+}({ \bm{k}})}{2T}-\tanh \dfrac{E_{-}({%
 \bm{k}})}{2T}}{4\sqrt{\varepsilon _{-}^{2}({ \bm{k}})+\Delta
^{2}}}.
\end{split}
\label{eq:omega_sum}
\end{equation}%
Finally, substituting this equality in (\ref{eq:dF}) we end up with the
function which expansion coefficients we will study further: 
\begin{equation}
\begin{split}
F^{\prime }& =-2T\sum_{{ \bm{k}}\omega }\frac{\Delta }{[\omega
+i\varepsilon _{+}({ \bm{k}})]^{2}+\varepsilon _{-}^{2}({ \bm{k%
}})+\Delta ^{2}}+\frac{4\Delta }{U} \\
& \simeq A\Delta +B\Delta ^{3}+\dots
\end{split}
\label{eq:dF_omega}
\end{equation}%
As it follows from Eqs. (\ref{eq:F_expansion})-(\ref{eq:dF_expansion}),
the expansion coefficient of the free-energy $F$ at $\Delta ^{n}$ is
obtained by dividing by $n$ the coefficient of the expansion of $%
F^{\prime }$. For the first two coefficients we obtain 
\begin{equation}
A=-2T\sum_{{ \bm{k}}\omega }\frac{1}{[\omega +i\varepsilon _{+}({%
 \bm{k}})]^{2}+\varepsilon _{-}^{2}({ \bm{k}})}+\frac{4}{U}
\label{A}
\end{equation}%
and 
\begin{equation}
B=2T\sum_{{ \bm{k}}\omega }\frac{1}{\left\{ [\omega +i\varepsilon _{+}(%
{ \bm{k}})]^{2}+\varepsilon _{-}^{2}({ \bm{k}})\right\} ^{2}}.
\label{B}
\end{equation}%
If $B>0$, the DW-metal phase transition is of the second order, and only the
first two coefficients $A$ and $B$ are sufficient for its description. If $%
B<0$, the phase transition may be of the first order, and next coefficients $%
C$ and even $D$ if $C\leq 0$ are required for its description.

The sum over $ \bm{k}$ in Eq. (\ref{eq:dF_omega}) for macroscopic
sample is equivalent to the integral 
\begin{equation}
	\sum_{\bm{k}}=
	2 \int_0^{4 k_\text{F}} \frac{\diff k_x}{2\pi} \int_{-\pi/b}^{\pi/b} \frac{\diff k_y}{2\pi}.
	\label{kInt}
\end{equation}
The factor 2 appears because of two Fermi-surface sheets at $k_{x}\approx
\pm k_\text{F}$, and we assumed quarter filling $\pi/a = 4 k_\text{F}$ \cite{Jerome2004}. Usually, for simplicity the integration limits over $k_{x}$ 
in Eq. (\ref{kInt}) are
taken to be infinite: $k_{x} \to (k_{x} - k_\text{F}) \in (-\infty, \infty)$, and the
resulting logarithmic divergence of the coefficient $A$ is regularized by
the definition of the transition temperature $T_\text{c0}$ (see Appendix \ref%
{sec:regularization}). However, in this case for the dispersion in Eqs. (\ref%
{dispersion1}) and (\ref{dispersion}) all the coefficients $A,B,C,D$ vanish at the
same point at $T=0$ and $t_{b}^{\prime }=t_{b}^{\prime \ast }$, which does not allow one to determine 
the type of this phase transition. Hence, below we consider both infinite and finite integration
limits as well as the case beyond the linear approximation when the spectrum
in $k_x$ direction is $\propto \cos(k_x a)$.

\section{Calculation of the Landau expansion coefficients for the density
wave and its phase diagram in quasi-1D metals}\label{sec:coefficients}

In this section we take the quasi-1D electron (\ref{dispersion1}) linearised
in the $k_{x}$ direction, and explicitly calculate the first coefficients of
the Landau expansion for the DW free energy given by Eqs. (\ref{eq:dF_omega}%
)-(\ref{B}). This helps us to analyze the type of DW-metal phase transition
in organic metals.

\subsection{Infinite limits of integration over $k_{x}$}

The integration of Eq. (\ref{eq:dF_omega}) over $k_x$ in infinite limits for
the linearized electron dispersion (\ref{dispersion1}) gives the well-known
result for the DW free energy: 
\begin{equation}
F^\prime = - \frac{2 T}{\hbar v_\text{F}} \sum_{\omega}\left\langle \frac{%
\Delta }{\sqrt{[\omega +i\varepsilon _{+}(k_{y})]^{2}+\Delta^{2}}}%
\right\rangle _{k_{y}} + \frac{4\Delta}{U}.  \label{eq:Delta_SC_inf}
\end{equation}%
The free energy expansion coefficients are obtained by expanding the r.h.s.
of this equation over $\Delta^{2}$: 
\begin{equation}
F^\prime \simeq  - \frac{2 T}{\hbar v_\text{F}} \sum_{\omega }\left\langle 
\frac{\Delta \sgn \omega}{\omega +i \varepsilon_+(%
 \bm{k})} - \frac{\Delta^3 \sgn \omega}{2[\omega +i
\varepsilon_+( \bm{k})]^3} \right\rangle _{k_{y}} + \frac{4\Delta}{U}.
\end{equation}
The coefficient at the linear term 
\begin{equation}  \label{eq:A_inf}
A_\infty = - \frac{2 T}{\hbar v_\text{F}} \sum_{\omega }\left\langle \frac{%
\sgn \omega}{\omega +i \varepsilon_+( \bm{k})}
\right\rangle _{k_{y}} + \frac{4}{U}
\end{equation}
is divergent when sum over $\omega$ is taken. The divergence is regularized
by introducing the critical temperature $T_\text{c0}$ at which the DW phase
transition occurs when $t^\prime_b = 0$, see Appendix \ref%
{sec:regularization}. As a results we obtain 
\begin{equation}
A_\infty = - \frac{2}{\pi\hbar v_\text{F}} \left[ \ln \frac{T_\text{c0}}{T} + \psi\left( 
\frac{1}{2} \right) -  \left \langle\Re \psi\left(\frac{1}{2} + \frac{i\varepsilon_+( 
 \bm{k})}{2 \pi T} \right) \right \rangle_{k_{y}} \right] ,
\label{Areg}
\end{equation}
where $\psi (x)=\diff/\diff x \ln \Gamma (x)$ is the digamma function.

The type of the phase transition is defined by the coefficient in front of
the $\sim \Delta ^{3}$ term in the expansion. This coefficient is 
\begin{equation}
\begin{split}
B_{\infty }& = \frac{T}{\hbar v_\text{F}} \sum_{\omega }\left\langle \frac{%
\sgn \omega }{(\omega +i\varepsilon _{+}( \bm{k%
}))^{3}}\right\rangle _{k_{y}} \\
& = - \frac{1}{8\pi^3 T^2 \hbar v_\text{F}}\left\langle \Re \psi ^{\prime \prime
}\left( \frac{1}{2}+\frac{i\varepsilon _{+}( \bm{k})}{2\pi T}\right)
\right\rangle _{k_{y}}.
\end{split}
\label{eq:b_inf}
\end{equation}%

The self-consistency equation (SCE) for the order parameter $\Delta $ is
obtained from the condition $F^{\prime }=0$ for the minimum of free energy.
This SCE for the DW state, even in the presence of both spin and charge
coupling constants and in a magnetic field, was derived previously \cite%
{Grigoriev2005} using the equations for the Green's function. The analytical
expression for free energy $F$ up to a constant factor can be obtained from
the integration of these SCE over $\Delta $, and the missing coefficient can
be obtained from the comparison of Eq. (\ref{eq:Delta_SC_inf}) above with
the Eq. (22) of Ref. \cite{Grigoriev2005}. The comparison of Eqs. (\ref{Areg}%
) and (\ref{eq:b_inf}) above with Eqs. (31) and (32) of Ref. \cite%
{Grigoriev2005} gives the relations $A_{\infty }=2(K_{\sigma
}^{(1)}-1) /U$ and $B_{\infty }=2K_{\sigma }^{(3)} /U$ %(\textbf{!! correct !!}) 
between the functions $K_{\sigma }^{(1)}$ and $K_{\sigma }^{(3)}$
derived in Ref. \cite{Grigoriev2005}.

The phase transition changes its order from second
to first when the coefficient $B$ becomes negative. The function (\ref{eq:b_inf}) 
generally decreases as the parameter $\varepsilon _{+}$
grows, but this decrease depends on the details of $\varepsilon _{+}$, 
in particular, on the ratio $\Delta _ \text{AO}/t_{b}^{\prime }$. 
Below we analyze the dependence $B(\varepsilon _{+})$ on the DW-metal transition line 
and the type of this phase transition in quasi-1D organic superconductors.
%Although it never truly changes sign, when the coefficient becomes infinitesimally small any perturbation is able to turn the second order phase transition into the first order one.

\subsection{Application to organic metals}

\begin{figure*}[tb]
	\centering
	\begin{tikzpicture}[every node/.style={inner sep=0,outer sep=0}]
		\node (picture) {\includegraphics{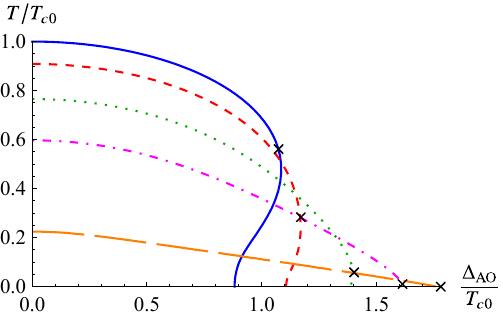}};
		\node[below=0.2cm,left=0.5cm] at (picture.north east) {(a)};
	\end{tikzpicture}
	\quad
	\begin{tikzpicture}[every node/.style={inner sep=0,outer sep=0}]
		\node (picture) {\includegraphics{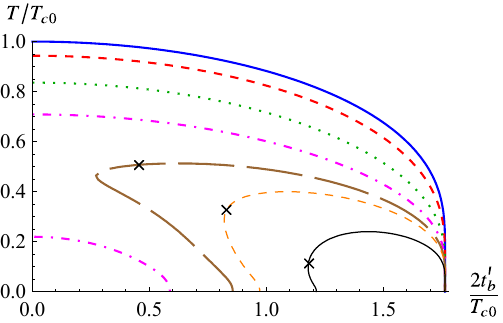}};
		\node[below=0.2cm,left=0.5cm] at (picture.north east) {(b)};
	\end{tikzpicture}
	
	\caption{Calculated transition lines as a function of 
		(a) anion ordering gap $\Delta_{ \text{AO}}$ for specific $t_{b}^{\prime }$:
		$t_{b}^{\prime }=0$ (solid blue line), $%
		2t_{b}^{\prime }=\Delta_{0}/2$ (dashed red line), 
		$2t_{b}^{\prime	}=0.75~\Delta _{0}$ (dotted green line), 
		$2t_{b}^{\prime }=0.9~\Delta _{0}$ (magenta dash-dotted line), and 
		$2t_{b}^{\prime }=0.99~\Delta _{0}$ (long-dashed orange line);
		(b) $t_{b}^{\prime }$ for specific anion ordering gap $\Delta_{ \text{AO}}$:
		$\Delta_ \text{AO}=0$ (solid blue line), 
		$\Delta_ \text{AO}=T_\text{c0}/2$ (dashed red line), 
		$\Delta_ \text{AO}=0.8~T_\text{c0}$  (dotted green line), 
		$\Delta_ \text{AO}=0.98~T_\text{c0}$ (dash-dotted magenta line), 
		$\Delta_ \text{AO}=1.09~T_\text{c0}$ (long-dashed brown line),
		$\Delta_ \text{AO}=1.15~T_\text{c0}$ (thin dashed orange line) and 
		$\Delta_ \text{AO}=1.3~T_\text{c0}$ (solid thin black line). 
		$T_\text{c}\left( \Delta_ \text{AO}\right) $ is	determined from the condition $A_{\infty }=0$ 
		and Eqs. (\ref{Areg}) and (\ref{epAO}). 
		The crosses on each line approximately indicate the points where the DW-metal phase 
		transition changes its type from second to first.   
		This phase diagram is qualitatively applicable to (TMTSF)$_{2}$ClO$_{4}$ and also 
		to (TMTSF)$_{2}$PF$_{6}$ at $\Delta_{ \text{AO}}=0$.
	}\label{FigPhDiaAO}
\end{figure*}

From Eqs. (\ref{tAO}) and (\ref{epm}) we obtain the antinesting term in
(TMTSF)$_{2}$ClO$_{4}$: 
\begin{equation}
%\varepsilon_+({ \bm{k}},{ \bm{k}}-{ \bm{Q}}_{0})\equiv 
\varepsilon_{+}({ \bm{k}})= -2t_{b}^{\prime }\cos
(2k_{y}b)+\Delta _ \text{AO}.  \label{epAO}
\end{equation}%
This antinesting term also describes (TMTSF)$_{2}$PF$_{6}$ as a limiting case $\Delta _ \text{AO}=0$. 

In the case of band-splitted electron dispersion, as in Eqs. (\ref{tAO}) for AO 
or (\ref{tZ}) for Zeeman splitting, for each band $\eta^{\prime} =\pm 1$ the sum over $\boldsymbol{k}$ 
in Eqs. (\ref{H})-(\ref{kInt}) also contains the sum over two bands $\eta =\pm 1$: 
$\sum_{\bm{k}}\to \sum_{\bm{k},\eta}$. In this sum only one of two terms, 
corresponding to $\eta^{\prime}\neq \eta$, contains the antinesting term (\ref{epAO}). 
The other perfect-nesting term, corresponding to $\eta^{\prime}= \eta$, adds a constant term 
(\ref{Tc0Reg}) to the coefficient $A$. It renormalizes 
the transition temperature $T_\text{c0}$, but does not change Eq. (\ref{Areg}) and 
the transition lines  normalized to $T_\text{c0}$ and shown in Fig. \ref{FigPhDiaAO}.
For different bands  $\eta^{\prime} =\pm 1$ the antinesting term has opposite sign, 
therefore the DW wave vector does not shift along $k_x$ for not very large band 
splitting (see Ref. \cite{Grigoriev2005} for the phase diagram at $t_{b}^{\prime }=0$). 
Since the digamma function 
$\psi\left( 1/2 + i\, x \right) $ is symmetric to $x\to -x$, both bands  
$\eta^{\prime} =\pm 1$ are described by the same formulas.

%In Eq. (\ref{epAO}) for (TMTSF)$_{2}$ClO$_{4}$ we may omit the term $\sim t_{b}^{\prime }$, assuming $t_{b}^{\prime }\ll \Delta _ \text{AO}$. This term $\sim t_{b}^{\prime }$ is important at $\Delta _ \text{AO}=0$ because it violates FS nesting and leads to the DW-metal phase transition with the increase of pressure, e.g., in (TMTSF)$_{2}$PF$_{6}$ and $\upalpha $-(BEDT-TTF)$_{2}$KHg(SCN)$_{4}$. 
%However, in (TMTSF)$_{2}$ClO$_{4}$ the violation of FS nesting is controlled mainly by $\Delta _ \text{AO}\gg t_{b}^{\prime }$, and $t_{b}^{\prime }$ only gives a small correction to the phase diagram by shifting it to a smaller $\Delta _ \text{AO}$ or to higher cooling rates. 
The DW-metal transition line $T_\text{c}(\Delta _ \text{AO},t_{b}^{\prime })$, given by
equation $A_{\infty }=0$ and shown in Fig. \ref{FigPhDiaAO}, suggests several 
interesting observations. Both $t_{b}^{\prime }$ and $\Delta _ \text{AO}$ suppress 
$T_\text{c}$, as expected. However, the increase of $t_{b}^{\prime }$ increases the 
critical value of $\Delta _ \text{AO}^{\ast}$ of the DW-metal phase transition. This 
counterintuitive observation can be shown analytically.
In the limit of zero temperature, one can obtain an exact expression for 
$\Delta_{ \text{AO}}^{\ast} (t_b')$. Using the asymptotic approximation for digamma 
function $\Re \psi( 1/2 +ix ) \simeq \ln x - 1 / (24x^2)$ at $x\gg 1$ and 
substituting Eq. (\ref{Areg}) to $A_{\infty }=0$, we get the equation for transition point
\begin{equation}
	\left\langle \ln \left( \varepsilon _{+}\right) \right\rangle
	_{k_{y}}=\ln (2\pi T_\text{c0})+\psi (1/2).
\end{equation}
Substituting Eq. (\ref{epAO}) and performing the integration over $k_{y}$ we
obtain the solution to this equation, valid at $\Delta _{\text{AO}}\leq
2t_{b}^{\prime }$:%
\begin{equation}
	\frac{\Delta _{\text{AO}}^{\ast }}{T_\text{c0}}=2\pi e^{\psi (1/2)}+\frac{%
		t_{b}^{\prime 2}/T_\text{c0}^{2}}{2\pi e^{\psi (1/2)}}.  \label{DAOm}
\end{equation}
In the limit $t_{b}^{\prime }=0$ we obtain $\Delta _{\text{AO}}/T_\text{c0}=2\pi
e^{\psi (1/2)}\approx 0.88$ (the intersection of the blue line with the $%
t_{b}^{\prime }$-axis in Fig. \ref{FigPhDiaAO}a). The maximal value of $%
\Delta _{\text{AO}}^{\ast }( t_{b}^{\prime }) $ is at $%
2t_{b}^{\prime }=\Delta _{\text{AO}}^{\ast }=2t_{b}^{\prime \ast }$: 
\begin{equation}
	\text{max}\left\{ \Delta _{\text{AO}}^{\ast }\right\} =2t_{b}^{\prime \ast
	}=4\pi e^{\psi (1/2)}T_\text{c0}\approx 1.764\,T_\text{c0} . \label{DAOmax}
\end{equation}%
In contrast to $\Delta _{\text{AO}}^{\ast }( t_{b}^{\prime }) $,
 within our model $t_{b}^{\prime \ast } $ is independent of $%
\Delta _{\text{AO}}$ as long as $\Delta _{\text{AO}}< \Delta _{\text{AO}}^{\ast }(
t_{b}^{\prime })$. This is illustrated in Fig. \ref{FigPhDiaAO}b, where
all transition lines cross at $t_{b}^{\prime }=t_{b}^{\prime \ast }$ at $T=0$. 

The second our result obtained for the antinesting term in Eq. (\ref{epAO}) 
is the dependence of $B_{\infty }$, given by Eq. (\ref{eq:b_inf}), on the parameters 
$\Delta _ \text{AO}$ and $t_{b}^{\prime }$. The point $B_{\infty }=0$ on the 
$T_\text{c}(\Delta _ \text{AO},t_{b}^{\prime })$ transition line, where this phase 
transition changes from the second to first order, is indicated by crosses in Fig. \ref{FigPhDiaAO}. 
From Fig. \ref{FigPhDiaAO}a we see that at any $t_{b}^{\prime }<t_{b}^{\prime \ast}$ 
there is a critical value $\Delta _ \text{AO}^{\rm{o}}$ of 
$\Delta _ \text{AO}^{\ast}(t_{b}^{\prime })$ such that 
at $\Delta _ \text{AO}>\Delta _ \text{AO}^{\rm{o}}$ there is a finite temperature 
interval $0<T_\text{c}<T_\text{c}^{\rm{o}}$, where $B_{\infty }<0$ and the DW-metal phase 
transition is of the first order. 

In contrast to this, the increase of $t_{b}^{\prime }$ on the interval 
$0<t_{b}^{\prime }<t_{b}^{\prime\ast}$ at fixed 
$\Delta _ \text{AO}<\Delta _ \text{AO}^{\rm{o}}$ does not leads to 
the sign change of $B_{\infty }$, given by Eqs. (\ref{eq:b_inf}) and (\ref{epAO}). 
This is illustrated in Fig. \ref{FigPhDiaAO}b by the absence of crosses 
on the transition lines at $\Delta _ \text{AO}<T_\text{c0}$. 
The dependence $B_{\infty }(t_{b}^{\prime })$ at $\Delta _ \text{AO}=0$ and $T\approx 0.6~T_\text{c0}$
is shown by the dashed green line in Fig. \ref{FigB}. This line crosses the abscissa axis at 
$t_{b}^{\prime }>t_{b}^{\prime\ast}$, and $B_{\infty }(t_{b}^{\prime })>0$ on the entire interval 
$0<t_{b}^{\prime }<t_{b}^{\prime\ast}$, which means the second-order DW-metal phase transition 
at any $t_{b}^{\prime }$. 

At $\Delta _ \text{AO}=0$ $B_{\infty }=0$ only at $t_{b}^{\prime }=t_{b}^{\prime\ast}$ and $T=0$, i.e. 
the interval $0<T_\text{c}<T_\text{c}^{\rm{o}}$ of the first-order phase transition is formally zero.
The point $t_{b}^{\prime }=t_{b}^{\prime\ast}$ and $T=0$ is very special: at this point not 
only $A_{\infty }=0$ and $B_{\infty }=0$, but also next coefficients of the Landau expansion vanish: 
$C_{\infty }=0$ and $D_{\infty }=0$. This degeneracy is a consequence of our oversimplified model, 
where we took the dispersion given by Eqs. (\ref{dispersion1}),(\ref{dispersion}),(\ref{epAO}) 
and the Fermi energy $E_F$ and the band width infinitely large, $\gg t_{b}^{\prime },\Delta _ \text{AO}$.
This degeneracy is lifted, for example, if one takes finite limits of integration over $k_{x}$, 
corresponding to a finite band width and Fermi energy, as described in the next subsection 
\ref{Sec:FiniteLimits}. A more realistic electron dispersion also gives a finite interval of the 
first-order DW-metal phase transition, as shown below in subsections \ref{Sec:4Harmonics} and 
\ref{Sec:NonLinearSpectrum}.  

\subsection{Finite limits of integration over $k_{x}$}\label{Sec:FiniteLimits}

Let us now take $\Delta _ \text{AO}=0$ and consider finite limits of integration $k_{x}\in [-K,K]$.
Then equation (\ref{eq:Delta_SC_inf}) becomes 
\begin{equation}
F^\prime -\frac{4 \Delta}{U}= - \frac{2 T}{\hbar v_\text{F}} \sum_{\omega }\left\langle \frac{%
\Delta \, f(\Delta )}{\sqrt{[\omega +i\varepsilon _{+}( \bm{k}%
)]^{2}+\Delta^{2}}}\right\rangle _{k_{y}}.  \label{eq:Delta_SC_K}
\end{equation}%
The additional factor $f(\Delta )$ is 
\begin{equation}
\begin{split}
f(\Delta )& =\frac{1}{\pi }\left[ \arctan \frac{\hbar v_\text{F}(K-k_\text{F})-2t_{b}\cos
(2k_{y}b)}{\sqrt{[\omega +i\varepsilon _{+}( \bm{k})]^{2}+\Delta ^{2}}}%
\right. \\
& -\left. \arctan \frac{-\hbar v_\text{F} k_\text{F}-2t_{b}\cos (2k_{y})}{\sqrt{[\omega
+i\varepsilon _{+}( \bm{k})]^{2}+\Delta ^{2}}}\right] .
\end{split}%
\end{equation}%
We expand the r.h.s of (\ref{eq:Delta_SC_K}) up to third order in $\Delta $: 
\begin{equation}
\begin{split}
& F^\prime -\frac{4 \Delta}{U} \simeq - \frac{2 T}{\hbar v_\text{F}} \sum_{\omega }\left\langle 
\frac{f(0) \sgn \omega }{\omega +i\varepsilon _{+}(%
 \bm{k})}\Delta \right. \\
& +\left. \frac{1}{2} \left[ \frac{f^{\prime \prime }(0) \sgn \omega }{\omega +i\varepsilon _{+}( \bm{k})} - \frac{f(0)%
\sgn \omega }{[\omega +i\varepsilon _{+}( \bm{k})]^{3}}\right] \Delta ^{3}\right\rangle _{k_{y}}.
\end{split}%
\end{equation}%
The coefficient at the linear term is 
\begin{equation}
A_{K}= - \frac{2 T}{\hbar v_\text{F}} \sum_{\omega }\left\langle \frac{f(0)%
\sgn \omega }{\omega +i\varepsilon _{+}( \bm{k}%
)}\right\rangle _{k_{y}} + \frac{4}{U}.
\end{equation}%
Due to finiteness of the limits of integration it is not divergent any more.
The coefficient at the qubic term is 
%\begin{equation}
%\begin{split}
%B_{K} &= \frac{2\pi T}{\hbar v_\text{F}} \sum_{\omega }\left\langle \frac{f(0)%
%\sgn\omega }{(\omega+i\varepsilon _{+}( \bm{k}%
%))^{3}} \right. - \\
%&\left. -\frac{f^{\prime \prime }(0)\sgn\omega }{%
%\omega +i\varepsilon _{+}( \bm{k})} \right\rangle_{k_{y}}.
%\end{split}
%\label{eq:b_K}
%\end{equation}
\begin{equation}
		B_{K} = \frac{T}{\hbar v_\text{F}} \sum_{\omega }\left\langle \frac{f(0)%
			\sgn\omega }{[\omega+i\varepsilon _{+}( \bm{k}%
			)]^{3}} -\frac{f^{\prime \prime }(0)\sgn\omega }{%
			\omega +i\varepsilon _{+}( \bm{k})} \right\rangle_{k_{y}}.
	\label{eq:b_K}
\end{equation}

\begin{figure}[tb]
	\centering\includegraphics{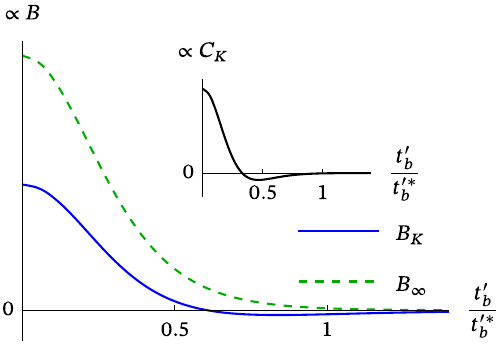}
	\caption{Schematic plots of expansion coefficients $B_{K}$ (solid blue line) 
		and $B_{\infty }$ (dashed green line) as
		functions of $t_{b}^{\prime }$ at $T \approx 0.35~T_\text{c0}$, 
		%The axis scales are arbitrary up to a constant factor.
		%The vertical axis scale is in units $4 
		%\protect\pi T_{c0}/\hbar v_\text{F}$, 
		where $T_\text{c0}$ is the critical temperature
		at $t^{\prime }_b = 0$, see Appendix \ref{sec:regularization}. The
		horizontal scale is in units $t_b^{\prime*}$ -- the value of $t_b^\prime$
		at which the DW-phase is destroyed, i.e. $A_K = 0$. 
		The inset shows similar	plot for the next expansion coefficient $C$ at $\Delta^6$. }
	\label{FigB}
\end{figure}

Numerically calculating the sum and the integral in Eq. (\ref{eq:b_K}) we confirm that 
$B_{K}$ at finite $K$ and low enough $T$ can indeed change its sign at $t^{\prime }_b < t^{\prime \ast}_b$.
Hence, the finite limits of integration change the order of DW-metal phase transition from second
to first one. The corresponding plot is presented in Fig. \ref{FigB}, 
the limits of integration are chosen as $K = \pi/a = 4 k_\text{F}$ in accordance with the quarter-filling, 
and for simplicity the following parameter values are taken: $\hbar v_\text{F} = t_b = 1$, 
$k_\text{F} = 1$, $U = 4$. The solid blue line $B_{K}(t^{\prime }_b)$ even at rather large temperature 
$T \approx 0.35~T_\text{c0}$ 
at certain parameter values crosses the abscissa axis at $t^{\prime \rm{o}}_b<t^{\prime \ast}_b$. %, 
%which indicates a finite temperature and pressure interval of the first-order DW-metal phase transition. 
This result predicts a finite temperature $0<T_\text{c}<T_\text{c}^{\rm{o}}$ and pressure interval of the 
first-order SDW-metal phase transition in (TMTSF)$_{2}$PF$_{6}$. 

\subsection{Influence of fourth harmonic in quasi-1D electron dispersion on the DW phase diagram}\label{Sec:4Harmonics} 

The electron spectrum given by Eq. (\ref{dispersion}) is oversimplified. Since all odd harmonics 
of the $k_y$ electron dispersion do not violate the FS nesting in our model, 
the simplest modification of electron dispersion in  Eq. (\ref{dispersion}) 
affecting the phase diagram is adding the fourth harmonics.  
When the limits of integration over $k_x$ are kept infinite but the fourth
harmonic $\propto \cos(4 k_y b)$ is introduced to the spectrum
the formal expression (\ref{eq:b_inf}) for the coefficient $B$ holds, but
the energy $\varepsilon_+( \bm{k})$ becomes 
\begin{equation}
\varepsilon_+( \bm{k}) = -2 t^\prime_b \cos(2 k_y b ) - 2 t^{\prime
\prime}_b \cos(4 k_y b).
\label{e4}
\end{equation}
This additional term reduces the coefficient $B$ and leads to dependence $B(t^\prime_b)$ similar to $%
B_K(t^\prime_b)$ (presented by solid line in Fig. \ref{FigB}), i.e. when
the limits of integration are finite. Hence, taking a more realistic electron spectrum by adding higher harmonics to Eq. (\ref{dispersion}) lifts the above mentioned degeneracy $A=B=0$ at $t^{\prime }_b =t^{\prime \ast}_b$ and $T=0$ even for infinite integration limits $K=\infty$. The fourth harmonic in quasi-1D electron dispersion decreases the coefficient $B$ of the Landau expansion for free energy and favors the first order of DW-metal phase transition.

\subsection{Nonlinearised spectrum}\label{Sec:NonLinearSpectrum}

We now discard the linear approximation of the spectrum in the $k_x$
direction, i.e. replace the spectrum (\ref{dispersion1}) with 
\begin{equation}
\varepsilon( \bm{k}) = -2 t_a \cos(k_x a) + t_\perp ( \bm{k}%
_\perp).
\label{eNL}
\end{equation}
In this case the analytical integration with respect to $k_x$ of (\ref%
{eq:dF_omega}) is not possible. %Also the integration limits are required to
%be finite: $k_x \in [-\pi/a, \pi/a]$.
%
% Accordingly we expand the r.h.s. of equation (\ref{eq:dF_omega}) in powers
% of $\Delta $: 
% \begin{equation}
% \begin{split}
% F^{\prime }& \simeq -2T\sum_{ \bm{k}\omega }\left( \frac{\Delta }{%
% (\omega +i\varepsilon _{+}( \bm{k}))^{2}+\varepsilon _{-}( \bm{%
% k})}\right. - \\
% & \left. -\frac{\Delta ^{3}}{[(\omega +i\varepsilon _{+}( \bm{k}%
% ))^{2}+\varepsilon _{-}( \bm{k})]^{2}}\right) -\frac{\Delta }{U}.
% \end{split}%
% \end{equation}%
Once again performing numerical integration over $k_{x}$, $k_{y}$ within finite limits 
and summation over $\omega$ we confirm that the coefficient $B$ at $\Delta ^{3}$
changes sign as function of $t_{b}^{\prime }$ at $t_{b}^{\prime }<t_{b}^{\prime \ast}$. 
Thus we conclude that the
first order phase transition emerges also when the linear approximation of
electron spectrum along $k_x$ is not made.

\section{Discussion and conclusion}\label{sec:discussion}

Our calculations show that a DW-metal transition, driven by an external parameter 
which deteriorates the FS nesting, typically goes by first order at low enough 
temperature. When such a driving parameter splits the energy spectrum and the 
Fermi surfaces, as in the case of anion ordering in (TMTSF)$_{2}$ClO$_{4}$ or 
of magnetic field acting via Zeeman splitting on a CDW, as described in 
Sec. \ref{sec:model}B, the temperature interval of the first-order phase 
transition is rather wide, $\sim T_\text{c0}/2$ (see Fig. \ref{FigPhDiaAO}a). 
This explains the spatial phase segregation and coexistence of SDW-metal or 
SDW-SC phases observed in (TMTSF)$_{2}$ClO$_{4}$. 
When such a driving parameter is pressure and the antinesting term in 
electron dispersion, as described by the amplitude $t_{b}^{\prime }$ of 
the second harmonic in Eq. (\ref{dispersion}), this interval is smaller 
and appears only when one goes beyond the simplest model by taking into 
account the finite bandwidth and Fermi energy (see Sec. \ref{Sec:FiniteLimits} 
and Fig. \ref{FigB}) or/and more realistic electron dispersion (see 
Secs. \ref{Sec:4Harmonics} and \ref{Sec:NonLinearSpectrum}). 

The model electron dispersions used in our calculations and given by 
Eqs. (\ref{dispersion1}),(\ref{dispersion}),(\ref{tAO}),(\ref{epAO}),(\ref{e4}),(\ref{eNL}) 
still differ from actual electron spectrum in organic metals (TMTSF)$_{2}$PF$_{6}$ 
and (TMTSF)$_{2}$ClO$_{4}$, e.g., calculated using DFT in Refs. 
\cite{Alemany2014,Guster_2020PF6,Guster2020}. The main difference comes from 
low triclinic crystal symmetry and higher harmonics of electron dispersion.  
Nevertheless, the result obtained about the negative coefficient $B$ of the Landau 
expansion for DW free energy and of the first-order phase transition between DW and 
metallic state at low $T$ are rather robust to small variations of electron 
dispersion and should survive also for more realistic electron spectrum.

The obtained first order of DW-metal phase transition explains the spatial 
segregation of SDW and SC/metal phases observed in (TMTSF)$_{2}$PF$_{6}$ and 
(TMTSF)$_{2}$ClO$_{4}$, or of CDW and SC/metal phases 
in $\upalpha $-(BEDT-TTF)$_{2}$KHg(SCN)$_{4}$, resulting to DW coexistence with 
superconductivity. The parameters and properties of phase nucleation during the 
first-order phase transition was extensively studied in various systems 
\cite{Umantsev2012,Kalikmanov2012Nucleation}. Nevertheless, a quantitative 
application of this theory to the particular DW--metal\slash SC phase transitions 
is still missing.

The phase segregation during the first-order DW-metal phase transition happens 
on a rather large length scale $\gtrsim 1$~\textmu m, which explains the observation 
of angular magnetoresistance oscillations (AMRO) in (TMTSF)$_{2}$PF$_{6}$ 
\cite{ChaikinPRL2014} 
and in (TMTSF)$_{2}$ClO$_{4}$ \cite{Gerasimenko2013}, i.e. 
the oscillating dependence of interlayer magnetoresistance on the 
tilt angle of magnetic field $\bm{B}$ \cite{Kartsovnik2004Nov}. 
The latter is possible only if the size of metallic islands along
 $x$-axis exceeds the so-called quasi-1D magnetic length 
 $l_B=\hbar /eBb \sim 1$~\textmu m \cite{Kartsovnik2004Nov,ChaikinPRL2014}. 
 In addition, the field-induced SDW (FISDW) \cite{Ishiguro1998,AndreiLebed2008-04-23}, 
 i.e. the oscillating dependence of the SDW transition temperature on the strength 
 $B$ of magnetic field, are also observed in (TMTSF)$_{2}$PF$_{6}$ 
 \cite{ChaikinPRL2014} and in (TMTSF)$_{2}$ClO$_{4}$ \cite{Gerasimenko2013}, 
 which is also possible only if the size of metal islands in SDW matrix exceeds 
 $l_B=\hbar /eBb \sim 1$~\textmu m. Most microscopic theories of SC-DW phase 
 segregation, including superconductivity inside soliton walls of DW order 
 parameter \cite{Kang2010,GGPRB2007,GrigPhysicaB2009}, 
 cannot explain such a large size of metal/SC domains, and, hence, contradict the 
 AMRO and FISDW experiments \cite{ChaikinPRL2014,Gerasimenko2013}. 
 A large size $d>1$~\textmu m of SC domains is also required 
 \cite{MagnetochemistryKochev2023} to explain the observed 
 \cite{Kang2010,ChaikinPRL2014,Gerasimenko2014} anisotropic SC transition by 
 the current percolation via SC domains in finite-size samples \cite{Kochev2021}.
Our scenario of DW-SC phase separation is also consistent with the severalfold 
enhancement of the upper critical field $H_\text{c2}$ observed in the coexistence 
phase of these organic superconductors \cite{Hc2Pressure,CDWSC}, because the SC 
domain size $\sim 1$~\textmu m is comparable to the magnetic penetration depth 
$\lambda$, which enhances $H_\text{c2}$ \cite{Tinkham}. In (TMTSF)$_{2}$ClO$_{4}$ 
the in-plane penetration depth is rather large \cite{Pratt2013}, 
$\lambda_{ab} ( T=0 ) \approx 0.86$~\textmu m, and increases even more at 
$T \to T_\text{cSC}$. Similarly large value of $\lambda $ is expected in 
(TMTSF)$_{2}$PF$_{6}$ and $\upalpha $-(BEDT-TTF)$_{2}$KHg(SCN)$_{4}$.

Above we have substantiated the first-order DW-metal phase transition by a direct 
calculation of the Landau expansion of the DW free energy for a generic quasi-1D 
electron dispersion, relevant to various organic superconductors. A different model 
of a multiband quasi-2D metal with several FS nested parts and also with additional 
reservoir states may also give a first-order phase transition and, hence, a spatial 
phase segregation \cite{Rakhmanov2020}. The latter is based on the non-monotonic 
dependence of chemical potential on pressure, obtained from the numerical solution 
of the system of nonlinear equations, derived using the minimization of the grand 
thermodynamic potential with respect to the DW order parameter $\Delta$ and the 
constraint of fixed electron number \cite{Rakhmanov2020}. This model is similar 
to that in Refs. \cite{Rakhmanov2013,Rice1970}. It is derived using several 
important assumptions \cite{Rakhmanov2020}, which may be relevant to iron-based 
superconductors \cite{Rakhmanov2013} or Cr \cite{Rice1970}, but are not directly 
applicable to the studied organic superconductors. The first assumption is an 
isotropic parabolic electron dispersion with equal masses and Fermi velocities 
of the electron and hole bands. In our case there is only one band, and the electron 
dispersion is strongly anisotropic and nonparabolic. As we have shown above, even 
much finer details of electron dispersion, such as the fourth harmonics of small 
amplitude, change the type of phase transition and, hence, strongly affect the phase 
segregation. Second, there is no intrinsic electron reservoir, which is also 
necessary for the main result of Ref. \cite{Rakhmanov2020}. The 
proposed \cite{Rakhmanov2020} appearance of small ungapped parts of the Fermi surface 
leads to a uniform superconductivity \cite{GGPRB2007,GrigorievPRB2008} with 
isotropic $T_\text{cSC}$, which contradict the experiments 
\cite{Kang2010,ChaikinPRL2014,Gerasimenko2014,Yonezawa2018} and  may not be 
energetically favorable in the non-uniform DW state. Moreover, the size of such 
ungapped FS pockets and the corresponding density of states would strongly depend 
on the DW order parameter $\Delta$, which should be accounted for during the variation 
of thermodynamic potential in Ref. \cite{Rakhmanov2020}. The third important 
assumption in Refs. \cite{Rice1970,Rakhmanov2013,Rakhmanov2020} is a commensurate 
SDW, while in organic superconductors the DW is incommensurate with crystal 
lattice. Small deviations of the optimal DW wave vector from a commensurate value 
often results to the commensurate-incommensurate first-order phase transition, e.g., 
as shown in Ref. \cite{Rice1970}. Fourth, the constraint of a fixed electron number, 
used in Refs. \cite{Rakhmanov2013,Rakhmanov2020}, assumes a constant electron density, 
which implies a fixed sample volume instead of a fixed pressure as in most experiments. Due to the different compressibility of DW and metallic phases, such a constraint may lead to a spatial segregation even in a simpler model of Ref. \cite{Vuletic}. Therefore, the model of Refs. \cite{Rice1970,Rakhmanov2013,Rakhmanov2020} cannot be directly applied to the studied quasi-1D organic superconductors, but its modification to such compounds would be interesting for further studies. 

In organic metals the DW phase transition temperature $T_\text{cDW}$ is usually 
much higher than the SC transition temperature $T_\text{cSC}$. This is a general 
situation, which happens in many other compounds where CDW or SDW coexists with 
superconductivity \cite{Review1Gabovich,ReviewGabovich2002}. This difference of 
$T_\text{cDW}$ and $T_\text{cSC}$ appears because for the formation of a DW the 
usually strong repulsive Coulomb $e$-$e$ interaction is suitable, 
while for the Cooper pairing only the $e$-$e$ attraction via phonons or via other 
mediating quasi-particle is needed, which usually competes with the strong Coulomb 
repulsion. The energy scale of SC is, normally, also much smaller than that of DW. 
Then there is no much difference between the DW-SC and DW-metal phase transition. 
Therefore, in our analysis we disregarded the influence of superconductivity on the 
DW phase diagram. Similar approximation to study the spatial inhomogeneity of SDW 
phase and the properties of superconductivity on a DW background was used in Refs. 
\cite{GGPRB2007,GrigorievPRB2008,GG,GrigPhysicaB2009}. However, when the DW is almost 
destroyed by the antinesting parameter, so that $T_\text{cDW}\sim T_\text{cSC}$, the 
SC may affect the DW. This mutual influence of SDW and SC can be analyzed by studying 
two order parameters simultaneously \cite{Imry1975,Watanabe1985,Ivanov2009}, but its 
realization for a more complicated DW phase diagram and with the antinesting term in 
electron dispersion is still missing. DW and SC compete, because both open a gap and 
require the electronic states on the Fermi level. Hence, the coefficient before the 
biquadratic term in the Landau free-energy expansion with two order parameters is 
positive. By analogy with a simpler case \cite{Watanabe1985}, this interplay is 
expected to reduce the DW region on the phase diagram and to favor the first-order 
phase transition between DW and SC. Hence, the emergence of superconductivity in the 
metallic domains below the SC transition temperature $T_\text{cSC}$ retains our main 
results and even expands the parametric region of the first-order phase transition 
and of spatial phase segregation.

\medskip
	 
To summarize, we investigated the density wave (DW) phase diagram and the type of 
DW-metal phase transition in layered organic superconductors. Even at zero 
temperature the DW is destroyed by an external parameter which violates the 
Fermi surface nesting and reduces the electronic susceptibility at the DW wave 
vector. This parameter may be the antinesting term in the electron dispersion, 
which can be controlled by external pressure as in (TMTSF)$_{2}$PF$_{6}$ or in 
$\upalpha $-(BEDT-TTF)$_{2}$KHg(SCN)$_{4}$. This antinesting parameter may also be 
the band splitting, e.g., as in (TMTSF)$_{2}$ClO$_{4}$ due to the anion ordering, 
controlled by the cooling rate. By the direct calculation of the Landau expansion 
coefficients of the density-wave free energy of quasi-1D metals with this antinesting 
parameter we have shown that the DW-metal phase transition at low temperature 
$T\ll T_\text{c0}$ is, usually, of the first order. This gives a microscopic 
substantiation of the DW/SC spatial phase segregation on a large length scale 
$\gtrsim 1$~\textmu m, indicated by the angular magnetoresistance oscillations or 
by field-induced SDW, observed in (TMTSF)$_{2}$PF$_{6}$ \cite{ChaikinPRL2014} and 
in (TMTSF)$_{2}$ClO$_{4}$ \cite{Gerasimenko2013}. More importantly, it explains 
unusual superconducting properties of organic metals in the coexistence phase. 
According to the model of conductivity anisotropy in heterogeneous superconductors 
\cite{Sinchenko2017,Grigoriev2017,Seidov2018,Kochev2021,KesharpuCrystals2021,
Grigoriev2023FeSe}, this phase segregation explains the anisotropic superconductivity 
onset observed in various organic superconductors \cite{ChaikinPRL2014,
Gerasimenko2014,CDWSC}. It is also consistent with a severalfold increase of the 
upper critical magnetic field $H_\text{c2}$ observed in (TMTSF)$_{2}$PF$_{6}$ 
\cite{Hc2Pressure} or $\upalpha $-(BEDT-TTF)$_{2}$KHg(SCN)$_{4}$ \cite{CDWSC} 
in the coexistence phase. The results obtained may also be relevant to many other 
superconductors, where superconductivity coexists with a charge- or spin-density wave.

\section{Acknowledgments}
%The work was carried out with financial support from the NUST "MISIS" grant No. K2-2022-025 in the framework of the federal academic leadership program Priority 2030. 
S.S.S. acknowledges the Foundation for the Advancement of Theoretical Physics and 
Mathematics ''Basis'' for grant \# 22-1-1-24-1. 
The work of V.D.K. was supported from the NUST "MISIS" grant No. K2-2022-025 in the 
framework of the federal academic leadership program Priority 2030. P.D.G. 
acknowledges the State assignment \# 0033-2019-0001 and the RFBR grants \# 21-52-12043 
and \# 21-52-12027. 

\appendix

\section{Electron dispersion with anion ordering in (TMTSF)$_{2}$ClO$_{4}$}\label{sec:AppA}

To substantiate Eq. (\ref{tAO}), we note that the wave vector $ \bm{Q}_{  \text{AO}}$ of AO potential $V_{  \text{AO}}$ connects the electron momenta $ \bm{k}$\ and $ \bm{%
	k}+ \bm{Q}_{  \text{AO}}$, similar to the CDW. The corresponding
Hamiltonian and energy spectrum is given by Eqs. (\ref{HMF})-(\ref{epm})
with the replacement $\Delta \rightarrow V_{  \text{AO}}$ and $ \bm{Q}%
\rightarrow  \bm{Q}_{  \text{AO}}$. As a result, the new dispersion is
given by Eq. (\ref{dispersion1}) with%
\begin{multline}
	t_{\perp }( \bm{k}_{\perp })=\pm \sqrt{V_{ \text{AO}%
		}^{2}+4t_{b}^{2}\cos ^{2}(k_{y}b)} \\
	= \pm \sqrt{V_{ \text{AO}}^{2}+2t_{b}^{2}%
	\left( 1+\cos (2k_{y}b)\right) },  \label{tAO1}
\end{multline}%
in agreement with Eqs. (A4) and (A5) of Ref. \cite{Alemany2014}. At $V_{\text{%
		AO}}^{2}+2t_{b}^{2}$ $\gg 2t_{b}^{2}$ one can expand the square root in Eq. (%
\ref{tAO1}), which gives%
\begin{equation}
	t_{\perp }( \bm{k}_{\perp })
	\simeq \pm \left[ \sqrt{V_{ \text{AO}}^{2}+2t_{b}^{2}}+\frac{t_{b}^{2}}{%
		\sqrt{V_{\text{AO}}^{2}+2t_{b}^{2}}} \cos(2k_{y}b) \right].
\end{equation}
This coincides with Eq. (\ref{tAO}), where 
$\Delta _{ \text{AO}}=\sqrt{V_{ \text{AO}}^{2}+2t_{b}^{2}}$
and $-2t_{b}^{\prime }=t_{b}^{2} / \sqrt{V_{  \text{AO}}^{2}+2t_{b}^{2}}$. 

In fact we can expand Eq. (\ref{tAO1}) into a Fourier series for $k_y$ and 
recover Eq. (\ref{tAO}) quite generally, getting 
$\Delta_\text{AO} = 4 \sqrt{4+x^2} E[4/(4+x^2)]t_b/\pi$,
$-2 t_b^\prime = 2 \sqrt{4+x^2} \! \left\{\! (2+x^2) E[4/(4+x^2)] 
- x^2 K[4/(4+x^2)] \!\right\} \! t_b/ (3 \pi)$,
$x=V_\text{AO}/t_b$, $K(m)$ and $E(m)$ is complete elliptic integrals of the first and second kind.
The ratio of the Fourier coefficient for the next $\cos(4 k_y b)$ harmonic to the coefficient for 
the $\cos{2 k_y b}$ harmonic is
\begin{equation}
	\frac{c_4}{c_2}(x) = - \frac{2}{5} \left[
		2 + x^2 - \frac{3}{2 + x^2 \left(1 - \frac{K\left[4/(4+x^2)\right] }{ E\left[4/(4+x^2)\right]} \right) }
	\right],
\end{equation}
which is equal to $-1/5$ at $x=0$ and monotonically asymptotically approaches 
zero as $\propto 1/x^2$. This justifies our use of Eq. (\ref{tAO}).
% In the opposite case $t_{b}^{2}$ $\gg \Delta _{ \text{AO}%
% }^{2}$ Eq. (\ref{tAO1}) gives%
% \begin{equation}
% 	t_{\perp }(\bm{k}_{\perp }) \approx 
% 	\pm  \sqrt{2} t_{b} \sqrt{
% 	1+\cos (2k_{y}b) } \left[ 1+ \frac{V_\text{AO}^2}{4 t_b^2 (1+\cos(2 k_y b))} \right] ,
% \end{equation}%
% % \begin{equation}
% % 	t_{\perp }( \bm{k}_{\perp })\approx \pm \left[ \sqrt{2}t_{b}\left(
% % 	1+\cos (2k_{y}b)\right) +V_{ \text{AO}}\left( k_{y}\right) \right] ,
% % 	\label{tAO4}
% % \end{equation}%
% which again coincides with Eq. (\ref{tAO}), where $2t_{b}^{\prime }=\sqrt{2}%
% t_{b}$ and $\Delta _{  \text{AO}}\left( k_{y}\right) $ is a slow monotonic
% function of $\left\vert k_{y}\right\vert $ on the interval $\left( 0,\pi
% /2b\right) $: $\Delta _{ \text{AO}}\approx \sqrt{2}t_{b}+V_{  \text{AO}}$ at $%
% k_{y}=\pm \pi /2b$, and $\Delta _{  \text{AO}}\approx \sqrt{2}t_{b}+V_{\text{AO%
% }}^{2}/4t_{b}$ at $k_{y}=0$. 

The strength of anion potential $V_{  \text{AO}}$
and of the corresponding band splitting energy $\Delta _{  \text{AO}}$ in
(TMTSF)$_{2}$ClO$_{4}$ is still debated. The early calculation in an
extended H\"{u}ckel-band model give the site-energy difference between two
independent TMTSF molecules about $V_{  \text{AO}}\approx 100$~meV \cite%
{Pevelen2001}, but the more recent DFT calculations suggest smaller value of
the anion ordering half gap $\Delta _{  \text{AO}}\approx 14$~meV \cite%
{Alemany2014}. For comparison, the estimated transfer integrals in (TMTSF)$%
_{2}$ClO$_{4}$ are\cite{Alemany2014} $t_{a}=263$~meV and $t_{b}=49$~meV. 

\section{Regularization of the divergence in coefficient $A_\infty$}\label{sec:regularization} 
The coefficient $A_\infty$, which is given by Eq.
(\ref{eq:A_inf}), is logarithmically divergent when the sum over $\omega$ is taken. This
divergence comes from the summation of $\sgn\omega/\omega$ term caused by the unrestricted 
integration over $k_x$. In real system they are limited by the first Brillouin zone 
and by finite bandwidth and Fermi energy. This divergent contribution can be 
regularized and expressed via the DW transition temperature $T_\text{c0}$ by
using the condition $A_\infty = 0$ at $T=T_\text{c}$.

Suppose that for $t_{b}^{\prime }=0$ the phase transition occurs at
temperature $T_\text{c0}$, where $\Delta =0$. Then from the condition $%
A_{\infty }=0$ it follows that 
\begin{equation}
\frac{2 T_\text{c0}}{\hbar v_\text{F}}\sum_{\omega }\left\langle \frac{%
\sgn\omega }{\omega }\right\rangle _{k_{y}}=\frac{4}{U}.
\label{Tc0Reg}
\end{equation}%
The divergent sum can be regularized by imposing a cutoff at $\omega \sim E_{F}$
which gives the equation for $T_\text{c0}$: 
\begin{equation}
\frac{U \nu_\text{F}}{2}  \ln \frac{2 e^{\gamma } E_{F}}{\pi T_\text{c0}}=1,
\end{equation}%
as in Ref. \cite{Solyom_Fundamentals_3}.
Here $\gamma \approx 0.577$ is the Euler constant.
Also the temperature $T_\text{c0}$ is related to the value of the gap $\Delta _{0}
$ at $T=0$ by usual BCS expression $\Delta_{0}\approx (\pi/e^\gamma) T_\text{c0}$.

Definition of $T_\text{c0}$ allows one to regularize the divergence. We subtract
and add the divergent term in the expression for $A_{\infty }$: 
\begin{equation}
\begin{split}
& A_{\infty }=-\frac{2 T}{\hbar v_\text{F}}\sum_{\omega }\left\langle \frac{%
\sgn\omega }{\omega +i\varepsilon _{+}( \bm{k}%
)}-\frac{\sgn\omega }{\omega }\right\rangle _{k_{y}} \\
& -\frac{2 T}{\hbar v_\text{F}}\sum_{\omega }\frac{\sgn\omega }{\omega }+\frac{4}{U}.
\end{split}%
\end{equation}
The first term is not divergent and is expressed via digamma function. For
the second term we perform the same regularization by introducing a cutoff
and find: 
\begin{equation}
\begin{split}
& -\frac{2 T}{\hbar v_\text{F}}\sum_{\omega }\frac{\sgn
\omega }{\omega }=-\frac{2}{\pi \hbar v_\text{F}}\ln \frac{2\alpha E_{F}}{\pi T} \\
%& =-\frac{4}{\hbar v_\text{F}}\left( \ln \frac{2\alpha E_{F}}{\pi }-\ln T+\ln T_\text{c0}-\ln T_{c0}\right) = \\
& =-\frac{2}{\pi \hbar v_\text{F}}\left( \ln \frac{2\alpha E_{F}}{\pi T_\text{c0}}+\ln 
\frac{T_\text{c0}}{T}\right) =-\frac{4}{U}-\frac{2}{\pi\hbar v_\text{F}}\ln \frac{T_\text{c0}}{T%
}.
\end{split}%
\end{equation}%
Substituting this back we finally obtain Eq. (\ref{Areg}), which is not
divergent any more.

\bibliographystyle{apsrev4-2}
\input{SDW.bbl}
%\bibliography{bib}

\end{document}